\def\e{{\rm e}}

\def\half{{1\over2}}
\def\cosb{\cos\beta}
\def\sinb{\sin\beta}

\def\abs#1{{\left|{#1}\right|}}

\font\seventeenbf =cmbx10 scaled\magstep3

\newcommand{\NPB}[3]{Nucl. Phys. {\bf B{#1}} (19{#2}) {#3}}

\newcommand{\PTP}[3]{Prog. Theor. Phys. {\bf {#1}} (19{#2}) {#3}}

\documentstyle[epsf,12pt]{article}
\makeatletter

\@addtoreset{equation}{section}
\makeatother
\begin{document}
\begin{titlepage}
\begin{flushright}
SAGA--HE--103\\
June 10,~1996
\end{flushright}
\vspace{24pt}
\centerline{\seventeenbf Explicit CP Breaking and Electroweak Baryogenesis}
\bigskip\bigskip
\centerline{Koichi Funakubo\footnote
{e-mail: \ funakubo@cc.saga-u.ac.jp},\ \ \  Akira Kakuto${}^*$\footnote
{e-mail: \ kakuto@fuk.kindai.ac.jp},}
\centerline{Shoichiro Otsuki${}^*$\footnote
{e-mail: \ otks1scp@mbox.nc.kyushu-u.ac.jp}
\ \ and\ \ \  Fumihiko Toyoda${}^*$\footnote
{e-mail: \ ftoyoda@fuk.kindai.ac.jp}}
\bigskip
\centerline{\it Department of Physics, Saga University,
Saga 840 Japan}
\centerline{${}^*$\it Department of Liberal Arts, Kinki University in
Kyushu, Iizuka 820 Japan}
\bigskip\bigskip
\abstract{
We investigate spatial behaviors of the  CP-violating angle $\theta$  by
solving the equation of motion of the two-Higgs-doublet model in the
presence of a small explicit CP breaking $\delta$. The moduli of the two Higgs
scalars are fixed to be the kink shape with a common width.
In addition to solutions  $\theta \sim O(\delta)$ in all the region,
we find several sets of two solutions of opposite signs, whose
magnitudes become as large as $O(1)$ around the surface of the bubble wall
while the CP violation in the broken phase limit is of $O(\delta)$.
Such set of solutions not only yield sufficient amount of the chiral charge
flux, but also avoid the cancellation in the net baryon number because
of a large discrepancy in their energy densities driven
nonperturbatively by the small $\delta$.}
\vfill\eject
\end{titlepage}
\baselineskip=18pt
\setcounter{page}{2}
\setcounter{footnote}{0}
%
%
\section{Introduction}
The electroweak baryogenesis\cite{reviewEB}
strongly depends on the  CP-violating angle $\theta(z)$ created at the
first-order  phase transition via bubble nucleation, where $z$ is the spatial
coordinate perpendicular to the `planar' bubble wall.
As is well known, one of difficulties of the electroweak baryogenesis
in the minimal standard model may be that
the explicit CP-breaking parameters of the Kobayashi-Maskawa scheme are too
small to generate the observed baryon asymmetry of the universe.\par
In a previous article    \cite{FKOTTc}\footnote{ Ref.\cite{FKOTTc}
is referred to as I.},
we  examined the behaviors of $\theta(z)$ that is spontaneously
generated in the
two-Higgs-doublet model, by assuming that the moduli
of the two neutral scalars,
$\rho_i(z) (i=1,2)$, have the kink shape of a common width.
One of the solutions we  presented has a remarkable spatial dependence such
that $\theta(z)$  becomes as large as
$O(1)$ around the surface of the bubble wall while it completely vanishes
deep in the broken and symmetric phase regions. Such a solution may be of
much importance in the electroweak baryogenesis since, as we
showed there, it does
yield a large amount of the chiral charge flux through the wall surface
at the  phase transition.\par
In this article we examine  $\theta(z)$ in the
two-Higgs-doublet model in the presence of an explicit CP breaking $\delta$
at the transition temperature $T_C$,
also by fixing the moduli to be the common kink shape. The magnitude of the
breaking  parameter $\delta$ may not be largely different from those of
the Kobayashi-Maskawa scheme even when finite-temperature corrections are
taken into account.
Then a naive guess would be that $\theta(z)$ remains of $O(\delta)$
in all the region between the broken and symmetric phase limits. Actually we
give  such solutions obtained analytically. On the other hand, we
find several solutions whose $\theta(z)$'s become
as large as $O(1)$ around the wall surface, as important
as the one in the case of spontaneous CP violation($\delta=0)$.\par
As pointed out by Comelli $et\, al.$\cite{Comelli},
the explicit CP breaking
may be necessary to avoid the complete cancellation in the net baryon number
 expected from the symmetry of the solution
$\theta(z) \longleftrightarrow -\theta(z)$ in the case of $\delta=0$.
In the presence of $\delta\neq 0$, the energy density of the bubble with
$\theta^+(z)$ close to $\theta^{\delta=0}(z)$   and that with
 $\theta^-(z)$ close to $-\theta^{\delta=0}(z)$   no more degenerate. We give
an estimate that the relative enhancement factor due to the energy
difference between the two kinds of bubbles could be as large as $O(10)$
even for $\delta \sim O(10^{-3})$. Such a large relative enhancement factor
 would favor the formation of one of the two kinds of bubbles and
 guarantee the baryon asymmetry of the universe. \par
     In Section 2 we introduce the breaking parameter $\delta$
into the standard two-Higgs-doublet potential, and give the equation of
motion for $\theta(z)$.  In Section 3, we discuss the boundary conditions
to be satisfied by $\theta(z)$. In Section 4 we show examples of
$\theta(z) \sim O(\delta)$ obtained analytically.
Several solutions of $\theta(z) \sim O(1)$ around the bubble wall are presented
in Section 5. In Section 6, the energy difference  and the relative
  enhancement factor are estimated. Section 7 is devoted to concluding remarks.
%
%
\section{Explicit CP Breaking and Equation for $\theta$}
In order to clarify essential roles played by $\delta$, we examine the problem
under the following simplified conditions:\par\noindent
\begin{enumerate}
\item[(1)] One breaking parameter $\delta$ is introduced into the standard
two-Higgs-doublet potential  as
$(m_3^2\e^{-i\delta}\Phi_1^\dagger\Phi_2 + h.c.)$ while
$\lambda_5,\lambda_6,\lambda_7 \in {\bf R}$ is assumed.
\item[(2)] The magnitude of $\delta$ at $T_C$ is  small enough.
\item[(3)] Let VEV's of the respective neutral components of $\Phi_i (i=1,2)$
be
$(1/\sqrt{2})\rho_i(z)\e^{i\theta_i(z)}$. The two moduli $\rho_i(z)$'s
are assumed to take the kink shape of a common width $1/a$:
\begin{equation}
 \rho_1(z)=v\cos\beta(1+\tanh(az))/2,\quad \rho_2(z)=v\sin\beta(1+\tanh(az))/2,
 \label{kink-shape}
\end{equation}
where $v\cos\beta$ and $v\sin\beta$ are VEV's
of $\Phi_1$ and $\Phi_2$ respectively in the broken phase limit.
\end{enumerate}
Here the same convention as I is used for the parameters in the
effective potential.
The condition (1) may not be specific since $\delta$ is induced
from the soft-SUSY-breaking parameters in the MSSM\cite{Comelli}.\par
Following I, we postulate the effective potential
$V_{eff}$, which is considered to  include the radiative and
finite-temperature corrections, as follows:
\begin{eqnarray}
 V_{eff}(\rho_1,\rho_2,\theta)&=&
 \half m_1^2\rho_1^2+\half m_2^2\rho_2^2 + m_3^2\rho_1\rho_2\cos(\delta+\theta)
 +{{\lambda_1}\over8}\rho_1^4+{{\lambda_2}\over8}\rho_2^4   \nonumber  \\
&+&{{\lambda_3-\lambda_4}\over4}\rho_1^2\rho_2^2
 - {{\lambda_5}\over4}\rho_1^2\rho_2^2\cos2\theta
 - \half(\lambda_6\rho_1^2+\lambda_7\rho_2^2)\rho_1\rho_2\cos\theta
    \label{eq:geneVeff}\nonumber\\
&-& \left(A\rho_1^3+B\rho_1^2\rho_2\cos\theta
            + C\rho_1\rho_2^2\cos\theta + D\rho_2^3 \right),
\end{eqnarray}
where $\theta\equiv\theta_1 -\theta_2$. Here the $\rho^3$ terms just above
are expected
to arise at finite temperatures so that the kink shape moduli
(\ref{kink-shape})
of the bubble wall are realized for $\theta(z)=0$ and $\delta=0$ at $T_C$.
Then several relations among the parameters in
(\ref{eq:geneVeff}) are required as given in I.\par
In terms of dimensionless coordinate $y\equiv (1-\tanh(az))/2$,
the equation of motion for $\theta(y)$ derived from
$V_{eff}(\rho_1=v \cos\beta (1-y),\rho_2=v \sin\beta (1-y),\theta(y))$ is:
\begin{eqnarray}
  & &y^2(1-y)^2{{d^2\theta(y)}\over{dy^2}}+y(1-y)(1-4y){{d\theta(y)}\over{dy}}
  \nonumber\\
 &=&b\sin(\delta+\theta(y))+\bigl[c(1-y)^2-e(1-y)\bigr]\sin\theta(y)
+ {d\over2}(1-y)^2\sin(2\theta(y)),
	\label{eq:eq-theta}
\end{eqnarray}
where
\begin{eqnarray}
  b&\equiv& -{{m_3^2}\over{4a^2\sinb\cosb}},   \nonumber\\
  c&\equiv& {{v^2}\over{32a^2}}(\lambda_1\cot^2\beta+\lambda_2\tan^2\beta
            +2(\lambda_3-\lambda_4-\lambda_5)) -
{1\over{2\sin^2\beta\cos^2\beta}}
      \nonumber \\
   &=&  {{v^2}\over{8a^2}}(\lambda_6\cot\beta + \lambda_7\tan\beta),
                  \label{eq:def-bcd}\\
  d&\equiv& {{\lambda_5 v^2}\over{4a^2}},   \nonumber   \\
  e&\equiv& {v\over{4a^2\sin^2\beta\cos^2\beta}}
             \left( A\cos^3\beta+D\sin^3\beta-{{4a^2}\over v} \right)
                   \nonumber\\
   &=& -{v\over{4a^2}}\left({B\over\sinb}+{C\over\cosb}\right). \nonumber
\end{eqnarray}
In addition, the requirement
 that $(\rho_1,\rho_2) = (0,0)$  and   $(\rho_1,\rho_2) =
(v \cos\beta,v \sin\beta)$ to be local minima of $V_{eff}$ with $\theta=0$
leads to inequalities among the parameters for $\delta=0$:
\begin{equation}
  b>-1,\quad b-2e+3c>-1+(\lambda_3-\lambda_4-\lambda_5)v^2/4a^2.
\end{equation}\par
Note that  the explicit CP violation $\delta \neq 0$  breaks the symmetry
$\theta(y) \longleftrightarrow -\theta(y)$ of (\ref{eq:eq-theta}), which is
allowed  in the case of  $\delta =0$.

%
%
\section{Boundary Conditions Satisfied by $\theta$}
{\bf Broken phase limit}\par\noindent
     Suppose that, at $y\sim 0$, $\theta(y)$ is given as
\begin{equation}
 \theta(y)=\theta_0+a_0 y^\nu+({\rm h.o.t.}(y))
   \qquad(\nu>0,a_0\not=0),
   \label{exp-theta-0}
\end{equation}
where $({\rm h.o.t.}(y))$ means (higher order terms of $y$).
Inserting this into (\ref{eq:eq-theta}), we have
\begin{eqnarray}
 y^\nu[\nu^2a_0+({\rm h.o.t.}(y))]&=&
    [W_0+W_1y+W_2y^2] \nonumber\\
 &+&y^\nu[W_3a_0+({\rm h.o.t.}(y))]\\
 &+&y^{2\nu}[(-a_0^2/2!)W_4+({\rm h.o.t.}(y))],
     \nonumber
      \label{y-limit}
\end{eqnarray}
where the $y^\nu$ terms on the right hand side come from
$\sin(\theta(y)-\theta_0)$
 and the $y^{2\nu}$ terms from $\cos(\theta(y)-\theta_0)-1$.\par
That $\nu>0$ requires
\begin{equation}
 W_0\equiv b\sin(\delta+\theta_0)+(c-e+d\cos\theta_0)\sin\theta_0=0,
                                    \label{W_0}
\end{equation}
from which
\begin{equation}
 \tan\theta_0=-\frac{b\sin\delta}
     {b\cos\delta+c-e+d\cos\theta_0}.
    \label{tan-theta0}
\end{equation}
Without loss of generality, let us restrict $\theta_0$
as $-\pi/2\le\theta_0<3\pi/2$.
Because of $b \propto m_3^2\neq 0$ necessary to introduce $\delta$,
 $\theta_0\neq 0,\pi$.
Making use of $(\ref{W_0})$
\begin{eqnarray}
   W_1&\equiv&(-2c+e-2d\cos\theta_0)\sin\theta_0
        =2b\sin(\delta+\theta_0)-e\sin\theta_0,   \\
   W_2&\equiv&(c+d\cos\theta_0)\sin\theta_0
        =-b\sin(\delta+\theta_0)+e\sin\theta_0,   \\
W_3&\equiv&b\cos(\delta+\theta_0)+(c-e)\cos\theta_0+d\cos(2\theta_0)\nonumber\\
        &=&-d\sin^2\theta_0-b(\sin\delta/\sin\theta_0),\\
   W_4&\equiv&b\sin(\delta+\theta_0)+(c-e)\sin\theta_0
    +4d\sin\theta_0\cos\theta_0\nonumber\\
        &=&3d\sin\theta_0\cos\theta_0.
\end{eqnarray}\par
If we take $\nu=1$ for illustration, $a_0$ is
determined as
\begin{equation}
 a_0=\frac{2b\sin(\delta+\theta_0)-e\sin\theta_0}
     {1+d\sin^2\theta_0+b(\sin\delta/\sin\theta_0)}
\label{a0}
\end{equation}
from $\nu^2a_0=W_1+W_3a_0$. When $\nu$ is not an integer, $a_0$ is not
determined from the lower order relations.\par\noindent
{\bf Symmetric phase limit}\par\noindent
     Suppose that, at $\zeta\equiv 1-y\sim 0$, $\theta(\zeta)$ is given as
\begin{equation}
 \theta(\zeta)=\theta_1+b_0 \zeta^\mu+({\rm h.o.t.}(\zeta))
   \qquad(\mu>0,b_0\not=0).
   \label{exp-theta-1}
\end{equation}
Inserting this into (\ref{eq:eq-theta}), we have
\begin{eqnarray}
 \zeta^\mu[\mu(\mu+2)b_0
       +({\rm h.o.t.}(\zeta))]&=&
    [U_0+U_1\zeta+U_2\zeta^2] \nonumber\\
 &+&\zeta^\mu[U_3b_0+({\rm h.o.t.}(\zeta))]\\
 &+&\zeta^{2\mu}[(-b_0^2/2!)U_4+({\rm h.o.t.}(\zeta))].
     \nonumber
      \label{zeta-limit}
\end{eqnarray}\par
That $\mu>0$ requires $ U_0\equiv b\sin(\delta+\theta_1)=0$, so that
\begin{equation}
 \theta_1=\ell\pi-\delta\qquad(\ell=0,\pm1,\pm2,\cdot\cdot\cdot\cdot).
\end{equation}
Making use of this,
\begin{eqnarray}
   U_1&\equiv&-e\sin\theta_1
        =(-1)^\ell e\sin\delta,   \\
   U_2&\equiv&(c+d\cos\theta_1)\sin\theta_1
        =-((-1)^\ell c+d\cos\delta)\sin\delta,   \\
   U_3&\equiv&b\cos(\delta+\theta_1)
        =(-1)^\ell b,   \\
   U_4&\equiv&U_0=0.
\end{eqnarray}\par
If we take $\mu=1$ for illustration, $b_0$ is
given from $\mu(\mu+2)b_0=U_1+U_3b_0$ as follows:     \par\noindent
(i) For $e\neq 0$ and $b\neq (-1)^\ell 3$,
\begin{equation}
 b_0=\frac{(-1)^\ell e\sin\delta}
     {3-(-1)^\ell b}.
   \label{b0}
\end{equation}
(ii) For $e= 0$,  $b= (-1)^\ell 3$ while $b_0$ is not specified
$a\,priori$.\par
Our task is to solve the nonlinear and inhomogeneous differential equation
(\ref{eq:eq-theta}) with the boundary conditions
of $\theta_0$ and $\theta_1$ (the two-point boundary value problem). We take
the
simplest boundary conditions that $\theta_0\sim O(\delta)$ and
$\theta_1=-\delta$ throughout, which will give the lowest energy
configurations.
%
%
\section{Examples of Solutions $\theta \sim O(\delta)$ }
Assuming that $\theta(y)\sim O(\delta)$ in the interval $y\in[0, 1]$, let us
linearize (\ref{eq:eq-theta}) as \cite{Cline}
\begin{eqnarray}
  & &y^2(1-y)^2{{d^2\theta(y)}\over{dy^2}}+y(1-y)(1-4y){{d\theta(y)}\over{dy}}
  \nonumber\\
 &=&\biggl[b+(c+d)(1-y)^2-e(1-y)\biggr]\theta(y)
   + b\delta.
	\label{eq:eq-linear}
\end{eqnarray}
Taking $\nu=1$ for illustration,
suppose, for a moment, to search for the solution as the initial value
problem: $\theta(y)$ starting from $\theta(0)=\theta_0$
with $\theta^\prime (0)=a_0$ has to hit against $\theta(1)=\theta_1$. A naive
guess would be that the
solution $\theta(y)$ should be of $O(\delta)$, since  $\theta_0$ by
(\ref{tan-theta0}),   $a_0$ by (\ref{a0}) and $\theta_1=-\delta$ are
all of $O(\delta)$ in general.
Actually, numbers of such solutions to the linearized equation
are obtained even analytically. We show a few examples.\par\noindent
{\bf Example 1.} We assume $\theta(y)=p_0+p_1 y$. Given $(b,d)$, a set of
algebraic linear equations for $(p_0,p_1)$ obtained from (\ref{eq:eq-linear})
are satisfied for $(c=4-d,
e=(7\pm\sqrt{1+16b^2})/2)$, giving $p_0=-(b/(b+c+d-e))\times\delta$ and
$p_1=-(4/(e-3)) p_0$. For this $\theta(y)$ to satisfy  the
boundary condition $\theta_1=-\delta$, the unique choice is $(b,e)=(3,0)$.
Namely, for $(b,c,d,e)=(3,4-d,d,0)$, we have a  linear solution
\begin{equation}
     \theta(y)=-((3+4y)/7)\times\delta.    \label{example-1}
\end{equation}
{\bf Example 2.} In the similar way, for $(b,c,d,e)=(3,10-d,d,0)$, we have a
 quadratic solution
\begin{equation}
     \theta(y)=-((3+5y+5y^2)/13)\times\delta.    \label{example-2}
\end{equation}\par
In the case of spontaneous CP violation, we obtained an almost linear
solution $\theta^{\delta=0}(y)$ shown in Fig.1 of I,  for which
$-\theta^{\delta=0}(y)$ is also
a solution. However, once $\delta\neq 0$ is given, $-\theta(y)$
of (\ref{example-1}) or (\ref{example-2}) is no more a solution.
Although we have no cancellation in the net baryon
number in this case, such solutions of $\theta(y)\sim O(\delta)$ would
contribute at most marginally to the baryon asymmetry\cite{Cline}.
%
%
\section{Solutions $\theta \sim O(1)$ around Bubble Wall}
Suppose that for $\delta=0$ we have a set of solutions
$\pm \theta^{\delta=0}(y)$ which are not $O(\delta)$ but $O(1)$.
For $|\delta|$ small enough, we could expect a solution
$\theta^+(y)$ close to $\theta^{\delta=0}(y)$ and another one
$\theta^-(y)$ close to $-\theta^{\delta=0}(y)$. The both $\theta^{\pm}(y)$
satisfy the same boundary conditions
 $\theta_0\sim O(\delta)$ and $\theta_1=-\delta$ but
 $\theta^-(y)\neq -\theta^+(y)$.
Before giving numerical solutions $\theta^{\pm}(y)$,
we show how such solutions of $O(1)$ are obtained from
$\pm \theta^{\delta=0}(y)$  for sufficiently small $\abs{\delta}$.
For definiteness, we take $\nu=\mu=1$.\par\noindent
{\bf Solutions $\pm \theta^{\delta=0}(y)$}\par\noindent
In Fig.1 we show three $\theta^{\delta=0}(y)$ satisfying the boundary
conditions
$\theta_0=\theta_1=0$ together with  the corresponding
parameters  $(b^{(0)},c^{(0)},d^{(0)},e^{(0)})$.
\begin{figure}
 \epsfxsize = 7cm
 \centerline{\epsfbox{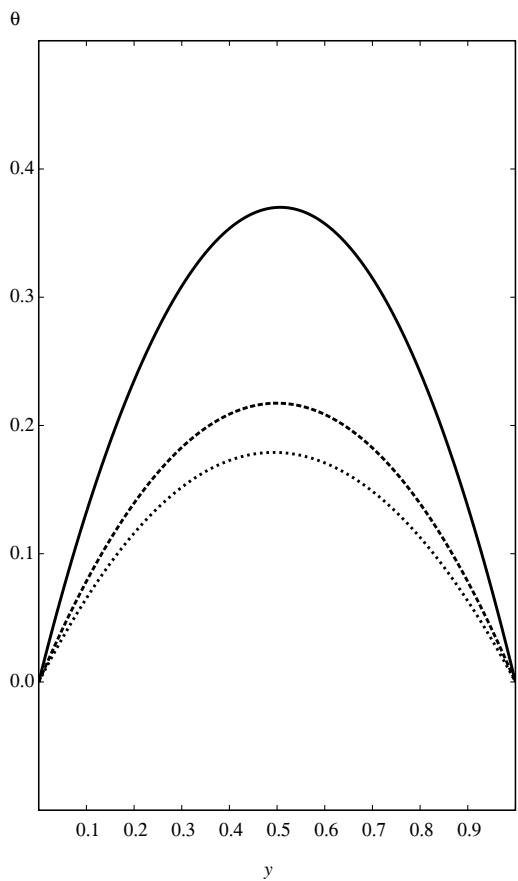}}
 \caption{Numerical solutions of $\theta^{\delta=0}(y)$ for
$(b^{(0)},c^{(0)},d^{(0)},e^{(0)})=(3,12.2,-2,12.2)$ (solid curve),
$(3,8.98,1,11.98)$ (dashed curve) and $(3,10,-0.2,11,8)$ (dotted curve).
The first is the one given in Fig.3 of I. For these, $\theta_0=\theta_1=0$.}
  \label{fig:1}
\end{figure}
One of them is the one given in Fig.3 of I.
Because $W_1=U_1=0$ for $\delta=0$, the parameters are required to satisfy
\begin{equation}
 b^{(0)}+c^{(0)}-e^{(0)}+d^{(0)}=\nu^2=1  \label{eq:del0-cond-1}
\end{equation}
from $(\nu^2-W_3|_{\delta=0})a_0^{\delta=0}=0$, and
\begin{equation}
 b^{(0)}=\mu(\mu+2)=3                    \label{eq:del0-cond-2}
\end{equation}
from $(\mu(\mu+2)-U_3|_{\delta=0})b_0^{\delta=0}=0$.
Of course, once we get a solution $\theta^{\delta=0}(y)$, its
$a_0^{\delta=0}$ and $b_0^{\delta=0}$ have been uniquely determined
to match the boundary conditions.
As is clear from the figure, $\theta^\prime(0)=a_0^{\delta=0}\sim O(1)$
and $\theta^\prime(1)=-b_0^{\delta=0}\sim O(1)$. These are what enable
$\theta(y\simeq0.5)$ to become as large as $O(1)$.\par\noindent
%
%
{\bf Solutions $\theta^{\pm}(y)$}\par\noindent
This suggests that we could have  solutions $\theta^{\pm}(y)\sim O(1)$ for
$\delta\not=0$ if $a_0\sim b_0\sim O(1)$. As shown in Section 3,
they are determined from the lower order relations when $\nu=\mu=1$.
Now we seek a set of parameters $(b,c,d,e)$, which incorporates
such solutions for $\nu=\mu=1$.\footnote{In practice, we are given
with the effective potential, so that $(b,c,d,e)$ are fixed. Then $\nu$ and
$\mu$ are determined by solving the equation for $\theta(y)$.
Since our purpose here is to show the possibility to have such solutions,
we trace an unusual course to find them.}\par
Put $b=b^{(0)}+\Delta b\times\delta$, $c=c^{(0)}+\Delta c\times\delta$,
$d=d^{(0)}$ and $e=e^{(0)}+\Delta e\times\delta$.
For sufficiently small $\abs{\delta}$, we have from (\ref{tan-theta0}), with
the use of (\ref{eq:del0-cond-1}),
\begin{equation}
 \theta_0\sim -\frac{b}{b+c-e+d}\times\delta\sim -b\times\delta,\quad i.e.,
\quad \delta/\theta_0 \sim -1/b.
\end{equation}
This implies that the denominator of $a_0$ in (\ref{a0})  is not
of $O(1)$ but $O(\delta)$.
Since the numerator of $a_0$ is $O(\delta)$, $a_0$ is now a quantity of
$O(1)$:
$a_0\sim b^{(0)}(2c^{(0)}-e^{(0)}+2d^{(0)})/(-\Delta b-\Delta c+\Delta e)$.
Because of (\ref{eq:del0-cond-2}), $b_0$ in (\ref{b0}) is also a
quantity of $O(1)$ for $e^{(0)}\neq 0$:  $b_0\sim -e^{(0)}/\Delta b$. By
suitably adjusting $(\Delta b,\Delta c,\Delta e)$
in such a way that $a_0 \sim a_0^{\delta=0}\sim O(1)$ and
$b_0 \sim b_0^{\delta=0}\sim O(1)$ match the boundary conditions,
we obtain a desired solution $\theta^+(y)$ close to $\theta^{\delta=0}(y)$.
For the same $(\Delta b,\Delta c,\Delta e)$ or $(b,c,d,e)$, we can find
another desired solution $\theta^-(y)$ close to
$-\theta^{\delta=0}(y)$.
In Fig.2 we show an example of $\theta^{\pm}(y)$.
\begin{figure}
 \epsfxsize = 7cm
 \centerline{\epsfbox{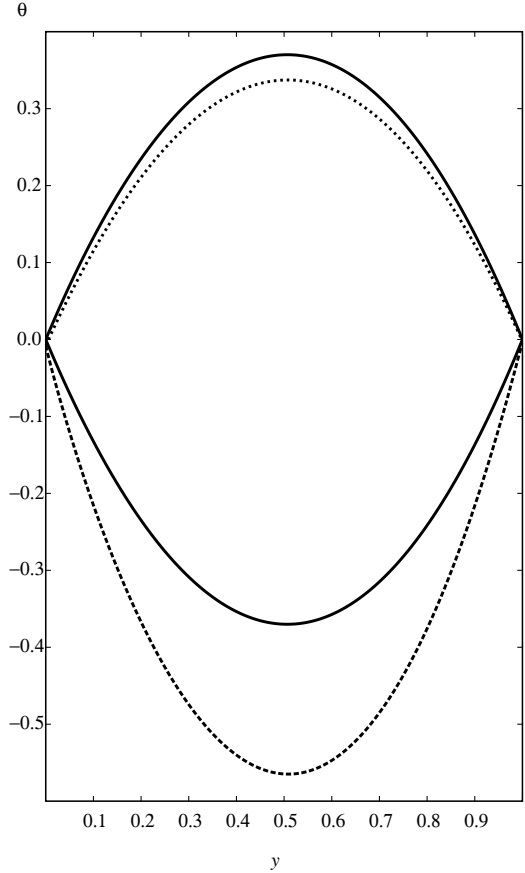}}
 \caption{Numerical solutions of $\theta^{\pm}(y)$ in which
$\delta>0$ is incorporated into the symmetric pairs of the solid-curve solution
$\theta^{\delta=0}(y)$ for $(b^{(0)},c^{(0)},d^{(0)},e^{(0)})=(3.12.2,-2,12.2)$
in Fig.~1 by the prescription explained in the text.
These pairs are given respectively by the upper and lower solid curves.
For $\delta=0.0025$ and $(b,c,d,e)=(2.98005,12.178375,-2,12.2)$,
$\theta^-(y)$  is given by the dashed  curve and $\theta^+(y)$ by the dotted
one. They have the common boundary values:
$\theta^{\pm}_0=-3.109066\times\delta
\sim -b^{(0)}\times\delta$ and  $\theta_1=-\delta$.}
 \label{fig:2}
\end{figure}
Note  that $-\theta^-(y)$ and $\theta^+(y)$ do not coincide with but
considerably differ from each other.  Note also that, while
$\theta_0 \sim -b\times\delta$ and $\theta_1=-\delta$,
$\theta^{\pm}(y)$ deviate nonperturbatively from
the corresponding $\pm\theta^{\delta=0}(y)$ in the intermediate region.
Presumably this may be due to the nonlinearity and the singular effects for
$\theta^{\prime\prime}(y)$ near $y=0,1$ of the differential equation
(\ref{eq:eq-theta}).\par
%
%
We can also find several solutions for other sets of the parameters
$(b,c,d,e)$, as long as they do not change the global structure
of the effective potential. Of course, they would not have
$\nu=\mu=1$ in general. The numerical
method is based on the relaxation algorithm. For sufficiently small $\delta$,
 say $\delta\in(0, 0.001]$, the both types of solutions
$\theta^{\pm}(y)$ can be found by starting from initial configurations
with opposite signs, respectively.
As we increase $\delta$, only $\theta^-(y)$ can be obtained starting from
{\it any} initial configuration. This would mean that for larger
$\delta$, $\theta^+(y)$ becomes more unstable or even does not exist
\footnote{This disparity in $\theta^{\pm}(y)$ would be triggered by
$\theta_0, \theta_1 <0$ for $\delta >0$ in our choice and amplified
nonperturbatively as remarked above.}.
Obviously the critical value of $\delta$, above which $\theta^+(y)$
is not found, depends on the parameters in the potential $(b,c,d,e)$.
Although its value is about $O(10^{-3})\sim O(10^{-2})$ in all the cases
we studied, we could not determine its definite value, since
it might depend on details in the algorithm, such as the convergence
parameter or initial configurations.
%
%
%
\section{Energy Density and Enhancement Factor}
The energy density of the wall per unit area is given by
\begin{eqnarray}
 {\cal E}&=&\int_{-\infty}^\infty dz \left\{
  \half\sum_{i=1,2}\left[\left({{d\rho_i}\over{dz}}\right)^2
    +\rho_i^2 \left({{d\theta_i}\over{dz}}\right)^2 \right]
   + V_{eff}(\rho_1,\rho_2,\theta) \right\}      \label{eq:energy-y}\\
 &=&
  \int_0^1 dy \left\{
  ay(1-y)\sum_{i=1,2}\left[\left({{d\rho_i}\over{dy}}\right)^2
           +\rho_i^2 \left({{d\theta_i}\over{dy}}\right)^2 \right]
  +{1\over{2ay(1-y)}}V_{eff}(\rho_1,\rho_2,\theta) \right\},
           \nonumber\\
   & \equiv &          av^2/3 +{\cal E}[\theta],
\end{eqnarray}
where the first term of the above line is the energy density of the
trivial solution $\theta^{\delta=0}(y)=0$. The second term is contributed
from a path $\theta(y)$ that connects the broken and symmetric  vacua.
For the trivial solution this term vanishes.\par
${\cal E}$ has two degenerate minima corresponding to
$\pm\theta^{\delta=0}(y)\not=0$.
 For the solid-curve solutions  in Figs.1 and 2, we have
\begin{equation}
  {\cal E}[\pm\theta^{\delta=0}] = -2.056\times10^{-3}av^2
          \sin^2\beta\cos^2\beta.
\end{equation}  \par
That $\delta\neq 0$ breaks the degeneracy. For  $\theta^{\pm}(y)$  shown in
\ref{fig:2},
in which  $\delta=0.0025$ is incorporated into these $\pm\theta^{\delta=0}(y)$,
we actually find that $0>{\cal E}[\theta^+]  >{\cal E}[\pm\theta^{\delta=0}] >
 {\cal E}[\theta^-]$, the energy difference being
\begin{equation}
  \Delta{\cal E}\equiv {\cal E}[\theta^-]-{\cal E}[\theta^+]
   =- 1.917\times10^{-2}av^2\sin^2\beta\cos^2\beta.
\end{equation}
This negative $\Delta{\cal E}$ means that the formation of the bubble
 with  $\theta^-(y)$ is favored over that with  $\theta^+(y)$.
The relative enhancement factor is given by
\begin{equation}
 \exp\left( -{{4\pi R_C^2\Delta{\cal E}}\over{T_C}}\right),
\end{equation}
where the radius of the critical bubble $R_C$ is approximately
given by $\sqrt{3F_C/(4\pi av^2)}$ with $F_C$ being the free energy of
the critical bubble.
Various authors estimate $F_C\sim (145\sim 160)T$.
If we take $F_C=145T$ and $\tan\beta=1$, the enhancement factor is
\begin{equation}
 \exp\left( -{{4\pi R_C^2\Delta{\cal E}}\over{T_C}}\right)
     =8.05.
\end{equation}
Such a large relative enhancement factor of $O(10)$
between the bubble with $\theta^-(y)$ and that with $\theta^+(y)$ would  surely
avoid the  cancellation in the chiral charge flux and guarantee the survival
of the net baryon number of the universe.\par
%
%
%
\section{Concluding Remarks}\par\noindent
Of course, the simplified condition (3) at the beginning of Section 2
that the wall moduli are fixed to be the kink shape is invalid for
the solutions $\theta(y)\sim O(1)$, and we have to solve a set of coupled
equations for $\theta(y)$ and $\rho_i(y)$ as done in \cite{FKOTb}.
As shown there, $\theta(y)$ remains to be $O(1)$
though its form is modified to certain extent while $\rho_i(y)$'s largely
deviate
from the kink shape.\footnote{For a numerical method how to obtain the chiral
transmission and reflection coefficients in such cases, see
\cite{FKOTa}.}
The kink-shape approximation is valid only for the solutions of $O(\delta)$
 as given in Section 4.
Though they have no counter partners as remarked there,
 the net baryon number would remain at most marginally  after the completion of
the phase transition.\par
On the other hand, as we gave an estimate in I for the solid-curve solution
$\theta^{\delta=0}(y)$ in Fig.1, such $\theta^{\pm}(y)$ that is able
to become $O(1)$ around the bubble wall could supply an efficient
chiral charge flux through the wall surface at the phase transition.
Note that in the broken
phase limit the CP violation is given by $\sin(\delta+\theta_0) \sim \delta$ at
$T_C$, so that, if $\abs{\delta}$ is small enough,
there should be no contradictions with the present experimental bounds.
Of course the CP violation completely vanishes in the symmetric phase
limit because of $\rho_i=0$ there.
That solutions with such features as presented here are allowed in a realistic
model would be highly significant in any scenario of the electroweak
baryogenesis, since there may be numbers of possible mechanisms to diminish
the net chiral charge or the net baryon number before the completion of the
phase transition.\par
Our estimate in Section 6 suggests an interesting possibility that
the effect of a small $\abs{\delta}$ is
nonperturbatively amplified to yield a large relative
enhancement factor to favor the formation of only one of the two kinds of
the bubbles, which are the symmetric partners in the absence of the explicit
CP violation.
Further our numerical analysis suggests that the bubble with higher
energy would be metastable or could not exist for larger $\abs{\delta}$.
If the possibility is realized, we could be free from the disgusting
cancellation in the baryon number of the universe.\par
\vspace{24pt}
This work was partially supported by Grant-in-Aid for Encouragement of
Young Scientist of the Ministry of Education, Science and Culture,
No.08740213 (K.F.).
%

%
%
\end{document}